\documentclass[aps,preprint,groupedaddress,showpacs,nofootinbib,amsmath,
amssymb]
{revtex4}

\usepackage{bm,feynmf,bbm}

\begin{document}

\title{Testing Spontaneous Parity Violation at the LHC}

\author{Chin-Aik Lee}
%\email{jlca@udel.edu}
\affiliation{Bartol Research Institute, Department of Physics \& Astronomy,
University of Delaware,
Newark, DE 19716}

\author{Qaisar Shafi}
%\email{shafi@bartol.udel.edu}
\affiliation{Bartol Research Institute, Department of Physics \& Astronomy,
University of Delaware,
Newark, DE 19716}

\date{\today}

\begin{abstract}
We construct a supersymmetric $SU(2)_L \times SU(2)_R \times
U(1)_{B-L}$ model 
in which a discrete symmetry ($C$-parity) implements strict 
left-right symmetry in the scalar (Higgs) sector. Although two 
electroweak bidoublets are introduced to accommodate the observed fermion 
masses and mixings, a natural missing partner mechanism insures that a 
single pair of MSSM Higgs doublets survives below the left-right symmetry
breaking scale. If this scale happens to lie in the TeV range, 
several new particles potentially much lighter than the $SU(2)_R$ charged gauge
bosons $W_R^{\pm}$ will be 
accessible at the LHC. 
\end{abstract}

\preprint{BA-07-23}


\maketitle

% \tableofcontents

Left-right symmetric models, in which the observed parity violation 
arises due to spontaneous symmetry breaking, have been extensively 
studied in the past \cite{lr,susylr,Dvali:1997uq}. Reference
\cite{Dvali:1997uq},
in particular, considered 
supersymmetric (SUSY) models based on the gauge group $SU(2)_L
\times SU(2)_R
\times 
U(1)_{B-L}$. It was shown in \cite{Dvali:1997uq} how the MSSM $\mu$ problem can
be
neatly 
resolved and hybrid inflation realized. For the most part, the 
discussion in \cite{Dvali:1997uq} assumed that the scalar (Higgs) sector
of the 
theory did not respect left-right symmetry. A brief discussion near the
end of the paper did, however, indicate that a left-right symmetric 
scalar sector may lead to several additional (non-MSSM) states in the low 
energy theory.
Our main goal in this paper is to provide a realistic SUSY
$SU(2)_L \times SU(2)_R \times U(1)_{B-L}$ model in which strict left-right
symmetry 
is enforced in all of the sectors by a discrete
$C$-parity \cite{Lazarides:1985my,Kibble:1982ae,Mohapatra:1979nn}.
($C$-parity
interchanges left and right and conjugates the representations. It also
requires $g_L=g_R$, where $g_{L,R}$ denotes the $SU(2)_{L,R}$ gauge couplings
respectively\footnote{It is well known that C-parity (also known as $D$-parity)
is contained
in SO(10)\cite{Lazarides:1985my,Kibble:1982ae,Mohapatra:1979nn}}.)
We will assume that the scale of spontaneous parity violation 
($M_R$) exceeds the SUSY breaking scale of around a TeV.
In order to accommodate the observed fermion masses and mixings one needs, it
turns out, two electroweak bidoublets. Fortunately, with manifest 
left-right symmetry in the scalar sector, a natural missing partner 
mechanism insures that with $M_R \gg M_W$, the low energy theory essentially 
coincides with the MSSM. However, if the scale of spontaneous parity 
violation $M_R$ happens to be in the multi-TeV range, the model predicts 
the existence of several new charged and neutral states which can be
significantly lighter than, say $W_R$, and therefore accessible at the 
LHC. In the simplest models the spontaneous breaking of
left-right 
symmetry can be achieved by employing $SU(2)_{L,R}$ doublet fields. Symmetry 
breaking with $SU(2)_{L,R}$ triplet fields is also possible and gives rise 
to additional states at low energy carrying two units of electric 
charge \cite{Chacko:1997cm,Dar:2005hm}.

% The Standard Model (SM) is a chiral gauge theory, meaning that it violates
% parity maximally. Na\"{i}vely, this strongly suggests that nature is both
% chiral
% and
% violates charge conjugation maximally. However, it is
% possible that nature so happens to be invariant under charge conjugation ($C$)
% at a
% fundamental level and the
% chiral nature of the SM that we currently observe is a consequence of a
% spontaneous symmetry breaking (SSB) of $C$. Left-right models with $C$-parity
% turn out to be one way of restoring this symmetry.
% \footnote{This symmetry is an
%  internal symmetry which does not transform spacetime nontrivially. It is a
% generalized
% charge conjugation which maps matter to
% antimatter and vice versa. However, it is
% not a true charge conjugation either because it is unitary and not
% antiunitary. In addition, it does not commute with the gauge symmetries.}
We extend, following \cite{lr},
the Standard Model (SM) electroweak gauge group $SU(2)_L \times U(1)_Y$ with its
chiral
matter content to a left-right gauge group $SU(2)_L \times
SU(2)_R \times U(1)_{B-L}$, where $B-L$ is baryon
number minus lepton number and
\begin{equation}
Y=B-L + T_{3R},
\end{equation}
where $T_{3R}$ is the third generator of $SU(2)_R$ in the Pauli basis.
$C$-parity shows up as a $\mathbb{Z}_2$
automorphism of the gauge group interchanging $SU(2)_L$ with $SU(2)_R$ and
conjugating $U(1)_{B-L}$ (and also color). In such a model, matter turns up
as the parity pairs $(\bm{3},\bm{2},\bm{1})_{\frac{1}{3}} \oplus
(\bm{\overline{3}},\bm{1},\bm{2})_{-\frac{1}{3}}$ and
$(\bm{1},\bm{2},\bm{1})_{-1} \oplus (\bm{1},\bm{1},\bm{2})_{1}$.

The electroweak symmetry breaking is implemented with a $C$-even bidoublet
$(\bm{1},\bm{2},\bm{2})_0$.
% \footnote{This is the case if we wish to get the
% matter Yukawa couplings directly from $H\psi \psi^c$. If we drop this
% condition, the electroweak Higgs need no longer come from an even bidoublet.
% This is the case if the matter couplings come from some nonrenormalizable
% coupling or we have additional massive vectorlike fermion fields which are
% indirectly responsible for the matter Yukawa couplings.}
When it acquires a nonzero vacuum expectation value
(VEV), it breaks both $SU(2)_L$ and $SU(2)_R$. However, this is not what we
observe experimentally; we have not observed any $SU(2)_R$ gauge bosons so far.
What this suggests is that $SU(2)_R \times U(1)_{B-L}$ -- but not $SU(2)_L$ --
is
broken to $U(1)_Y$ at some energy scale higher than the electroweak scale --
which might be the Grand Unified (GUT) scale or some other intermediate scale --
and thereby breaking $C$-parity as well. If a Higgs field is
responsible
for this symmetry breaking, then it has to be a singlet under $SU(3)_C$ and
$SU(2)_L$ but not under $SU(2)_R$. 

The simplest choice for breaking $SU(2)_R \times U(1)_{B-L}$ down to $U(1)_Y$
would be an $SU(2)_R$
doublet $\Phi_R$. Later, we will also analyze $SU(2)_R$ triplets $\Delta_R$.
Because of $C$-parity, we also need to include an $SU(2)_L$
doublet $\Phi_L$ as well. In a supersymmetric
model, to cancel the gauge anomaly and to be able to construct simple
superpotentials which can give rise to the Higgs mechanism, we also need to
introduce the complex conjugates of these fields, $\Phi_L^c$ and $\Phi_R^c$ to
the model. In addition, the conjugate fields are needed so that the $SU(2)_R
\times U(1)_{B-L}$ breaking will not lead to some nonzero $D$-terms, thereby
breaking SUSY at too high a scale. To summarize, it would seem that a minimal
left-right model with
$C$-parity
requires at least
three generations of matter, a Higgs bidoublet $H$ and possibly $\Phi_L$ and
$\Phi_R$ (or left and right triplets) plus their $C$-parity conjugates. 

Now from the fact that the $\Phi_R$'s and/or $\Delta_R$'s acquire nonzero VEVs
but the $\Phi_L$'s and the $\Delta_L$'s do not, we see that the VEV of some
superfields are not the same as their $C$-conjugates. This means that
$C$-parity is spontaneously broken at the same scale as the $SU(2)_R$ breaking
scale.
% The supersymmetric version of this model
% is free of gauge anomalies; the matter sector does not contribute any
% anomalies
% and neither does the Higgs sector.

% \subsection{Overview}
% We will analyze various supersymmetric left-right models with $C$-parity. In
% some of these models, we have unwanted
% left-right mass relations which may have to be broken by nonrenormalizable
% couplings whereas in others, like two Higgs bidoublet models, we can have MSSM
% at low energy scales with no left-right mass relations. This latter model is
% the most promising. In particular, previous analyses like those of Refs.
% \cite{Magg:1979tz,Dvali:1997uq} only considered the part of the Higgs sector
% consisting of $\Phi_L$ and $\Phi_R$ while neglecting the electroweak
% bidoublet.
% Those analyses are misleading because realistic left-right models contain
% Higgs
% bidoublets and these bidoublets can interact with $\Phi_L$ and $\Phi_R$. In
% particular, we will demonstrate in this article that their conclusions can
% easily be avoided simply by introducing the Higgs bidoublet.

% \section{Our main model}
The minimal
matter content consists of three generations of $L
(\bm{1},\bm{2},\bm{1})_{-1}$, $L^c (\bm{1},\bm{1},\bm{2})_{1}$, $Q
(\bm{3},\bm{2},\bm{1})_{\frac{1}{3}}$ and $Q^c
(\bm{\overline{3}},\bm{1},\bm{2})_{-\frac{1}{3}}$. A Yukawa coupling like $L H
L^c$ or $Q H Q^c$ predicts that $Y^U=Y^D$ and $Y^E=Y^D$. This is a typical
left-right prediction. However, we know that this is definitely not the case for
all the
three generations for the case of the quarks. One way around
this
relation is to double the number of bidoublets and make their electroweak VEVs
point in different directions (i.e. the ratios $\langle H_{ui} \rangle/\langle
H_{di} \rangle$ are different) and write down couplings to both of these
bidoublets. We now need to explain why the VEVs point in different directions.
Let us call the bidoublets $H_i$, $i=1,2$. 
The terms responsible for the electroweak scale VEVs are the radiative
corrections coming primarily from couplings to the third generation after
SUSY is broken. To get MSSM at low energies, we only want a pair of $SU(2)_L$
doublets to survive at low energies. This can be achieved if say only the $H_u$
from $H_1$
and $H_d$ from $H_2$ remain light. This can be achieved via the ``missing
partner
mechanism'' \cite{mpSU5}.
Let us say we have the couplings $\Phi_L H_1 \Phi_R$ and $\Phi_L^c H_2
\Phi_R^c$.
Then, after both $\Phi_R^c$ and $\Phi_R$ acquire $M$-scale VEVs, $\Phi_L$ pairs
up
with $H_{d1}$ and $\Phi_L^c$ pairs up with $H_{u2}$. The remaining light
superfields are none other than $H_{u1}$ and $H_{d2}$. With such a mechanism, we
can break the unwanted relation $Y^U=Y^D$ \footnote{
% We would also like to
% point out that there is another alternative to using the missing partner
% mechanism to explain why $Y^U\neq Y^D$. If we only have one Higgs bidoublet
% $H$ but also introduce an $SU(2)_R$ Higgs triplet $(\bm{1},\bm{1},\bm{3})_0$
% (as well as its $C$-conjugate $(\bm{1},\bm{3},\bm{1})_0$) and include the
% superpotential couplings $(\bm{1},\bm{1},\bm{3})_0 \Phi_R^c \Phi_R$ and
% $(\bm{1},\bm{1},\bm{3})_0^2$ (plus their $C$-conjugates), then the $SU(2)_R$
% triplet can get a nonzero induced VEV and with the nonrenormalizable coupling
% $y' H \langle (\bm{1},\bm{1},\bm{3})_0 \rangle Q^c Q/\Lambda$ and the direct
% coupling $y H Q^c Q$. Then, $Y^U=y+y'c/\Lambda$ and $Y^D=y-y'c/\Lambda$ where
% $c$ is the VEV of the $SU(2)_R$ Higgs triplet. If $\Lambda$ is around the
% Planck scale and $M$ is around the GUT scale, then the up-down relation can be
% modified appreciably for the first and second generations but not the third.
Alternatively, with a single bidoublet one could introduce the superfields
$(\bm{1},\bm{3},\bm{1})_0$ and $(\bm{1},\bm{1},\bm{3})_0$, and employ
nonrenormalizable couplings involving the VEV $\langle (\bm{1},\bm{1},\bm{3})_0
\rangle$ to break the relation $Y^U=Y^D$.
}.
% \begin{figure}
% \[
% \begin{array}{c|c|c|c|c||c}
% \Phi_L 	& 	& H' 	& 	& \Phi_L^c 	& H\\
% \hline
%  	& 	& H'_{u} & \longleftrightarrow	& \Phi_L^c 	& H_u\\
% \Phi_L 	& \longleftrightarrow	& H'_{d} & 	& 		& H_d
% \end{array}
% \]
% \caption{Pairing up $\Phi_L$ with $H'$; every superfield gets paired up}
% \label{fig:full_pairing}
% \end{figure}
% 
% \begin{figure}
% \[
% \begin{array}{c|c|c|c|c|c}
% \Phi_L 	& 	& H_1 		& H_2 		& 	& \Phi_L^c\\
% \hline
%  	& 	& H_{u1} 	& H_{u2} 	& \longleftrightarrow	&
% \Phi_L^c\\
% \Phi_L 	& \longleftrightarrow	& H_{d1} 	& H_{d2} 	& 
% & 
% \end{array}
% \]
% \caption{The ``missing partner mechanism''}
% \label{fig:mpm}
% \end{figure}

% \[
% \begin{array}{cccccc}
% \Phi_L 	& 	& H_1 	& H_2 	& 	& \Phi_L^c\\
% 	& 	& 
% \end{array}
% 
% \]

We note that even if we have the more general coupling
\begin{equation}
W \supset c_{11}\Phi_L H_1 \Phi_R + c_{12} \Phi_L H_2 \Phi_R + c_{21} \Phi_L^c
H_1 \Phi_R^c  + c_{22} \Phi_L^c H_2 \Phi_R^c,
\end{equation}
with the coefficients free to vary, generically, unless
$c_{11}c_{22}=c_{12}c_{21}$, we will still end up with the missing partner
mechanism.
The key here is to forbid the potentially huge $H_1H_2$ coupling, which is
taken care of
automatically if we have $\mathcal{R}$-symmetry. Without an
$\mathcal{R}$-symmetry, we need to come up with another mechanism to forbid
such couplings together with nonrenormalizable couplings like $(\Phi_R^c
\Phi_R)^n H_1 H_2/\Lambda^{2n-1}$, where $\Lambda$ denotes some cutoff scale.
% In other words, we do not need any necessarily special symmetry to get the
% desired superpotential.
%  However, it certainly is very hard to explain why the
% ratio  $Y^U/Y^D$ varies so much between the three generations in this more
% general scheme. This might suggest $c_{12}=c_{21}=0$. 

We may consider couplings like 
\begin{equation}
(L^c\Phi_R)(\Phi_R^c H L)/M^2,(L^c\Phi_R^c)(\Phi_R H L)/M^2,
(Q^c\Phi_R)(\Phi_R^c H Q)/M^2,(Q^c\Phi_R^c)(\Phi_R H Q)/M^2
\end{equation}
and their $C$-conjugates to break the left-right relation. These
nonrenormalizable couplings can be
gotten if we integrate over some massive fields with masses of order $M$, for
instance. However, in a minimal model, these couplings will be suppressed by
the Planck scale, which might lead to couplings which are too small.

% To get the seesaw mechanism, we can introduce the coupling $(\Phi_R
% L^c)^2/\Lambda$. Unfortunately, this coupling does not have an
% $\mathcal{R}$-charge of 2. One way to get this nonrenormalizable coupling is
% to
% introduce a singlet chiral superfield $N$ and introduce the superpotential
% couplings $N'^2$ and $N'L^c\Phi_R$. This gives us the double seesaw mechanism
% after we integrate over $N'$.

% Alternatively, we may simply settle for a $(H L)(H L)/\Lambda$
% coupling directly. No!!! $B-L$!!!

Next, let us try to construct the Higgs sector step by step. Our first
task is to come up with some mechanism to give the $\Phi_R$'s a nonzero VEV and 
one of the simplest ways of doing that is to introduce another $C$-even singlet
superfield $S$ and write down the
following superpotential\cite{Dvali:1994ms}\cite{Dvali:1997uq,Copeland:1994vg}:
\begin{equation}
W = \kappa S (\Phi_L^c \Phi_L + \Phi_R^c \Phi_R - M^2).
\label{eq:W} 
\end{equation}
The singlet $S$ plays an important role in hybrid inflation models
\cite{Dvali:1994ms}\cite{Dvali:1997uq,Copeland:1994vg}. (For a recent analysis
and additional references, see \cite{ur Rehman:2006hu}).
Without any loss of generality, we may redefine the phase of $S$ and the
$\Phi$'s so
that both $\kappa$ and $M$ are real and positive\cite{Dvali:1997uq}. 
We first note that in the absence of any gauge couplings, a generic
superpotential of the form $Sf(\Phi)$ will consist mostly of moduli. The
generic supersymmetric solution is $S=0$ and $f=0$. The latter condition singles
out a complex
submanifold of codimension 1 in $\Phi$ space. This moduli is not due to any
symmetry, accidental or otherwise, in general but to holomorphy and the form of
the superpotential. It may happen that in some cases, as in Eq. \ref{eq:W}, that
this result can also be reinterpreted in terms of an accidental $U(4)$
symmetry. 
% In particular, we will still have exactly the same number
% of moduli even if we add nonrenormalizable terms which explicitly break
% $U(4)$.
% In other words, nonrenormalizable terms will not change anything qualitatively
% or even quantitatively by all that much since their corrections at low
% energies
% will be suppressed by the cutoff scale.

Since we only want SUSY to be broken at the TeV scale, to a good approximation,
we can assume that the $D$ terms are zero at the scale of $M$.
This forces $\phi_L$ and $\phi_L^c$, and $\phi_R$ and $\phi_R^c$ to
acquire the same (complex conjugated) VEVs. We can always perform a gauge
transformation so that these VEVs are identical and real.
Since $SU(2)_L
\times U(1)_Y$ is only broken at the electroweak scale, we need $\langle
\phi_L\rangle=0$. 

Because of the moduli, the model as it currently stands contains some
additional light fields. Because of this,
we will not analyze it any further but will look into a more realistic
model. As mentioned in the introduction, a realistic minimal model will likely
contain \emph{two} electroweak doublets with a missing partner mechanism. In
addition, this pairing mechanism gets rid of precisely the light particles that
\cite{Dvali:1997uq} had noticed. Because of this, we can get MSSM at
low energies without the undesirable left-right mass relations. 
% In addition,
% in 
% models of the
% sort given in Ref \cite{Dvali:1997uq}, we need some fine-tuning of the soft
% SUSY
% terms in
% order to
% stabilize the moduli at $\langle \Phi_L \rangle=0$.
% However, we have to come up
% with acceptable soft SUSY terms.

\begin{table}
\[
\begin{array}{|c|c|c|c|}
\hline
\text{superfield} 	& \text{representation} 	& \text{superfield} &
\text{representation}\\
\hline
S 			& (\bm{1},\bm{1},\bm{1})_{0} 	&&\\
\Phi_L 			& (\bm{1},\bm{2},\bm{1})_{1} 	&
\Phi_R 			& (\bm{1},\bm{1},\bm{2})_{-1}\\
\Phi_L^c 		& (\bm{1},\bm{2},\bm{1})_{-1} 	&
\Phi_R^c 		& (\bm{1},\bm{1},\bm{2})_{1}\\
H_1 			& (\bm{1},\bm{2},\bm{2})_0 &&\\
H_2 			& (\bm{1},\bm{2},\bm{2})_0 &&\\
Q_i 			& (\bm{3},\bm{2},\bm{1})_{\frac{1}{3}} &
Q^c_i 			&
(\bm{\overline{3}},\bm{1},\bm{2})_{-\frac{1}{3}}\\
L_i 			& (\bm{1},\bm{2},\bm{1})_{-1} 	&
L^c_i 			& (\bm{1},\bm{1},\bm{2})_1\\
\hline
\end{array}
\]
\caption{The chiral superfields in the minimal model.}
\label{tab:fields}
\end{table}

Our model can be briefly summarized by the chiral superfield content of Table
\ref{tab:fields} and the superpotential
\begin{align}
W &= \kappa S(\Phi_L^c \Phi_L + \Phi_R^c \Phi_R - M^2) + \alpha \Phi_L H_1
\Phi_R
+ \beta \Phi_L^c H_2 \Phi_R^c +\nonumber\\
&\qquad Y^U Q H_1 Q^c + Y^D Q H_2 Q^c + Y^E L
H_2 L^c +
Y^{Dirac} L H_1 L^c + \frac{1}{\Lambda}\left( \Phi_R \Phi_R L^c L^c +
\Phi_L \Phi_L LL\right).
\label{eq:Wall}
\end{align}
(This superpotential respects $\mathbb{Z}_2$ matter parity.)
Note the appearance of nonrenormalizable terms needed to provide Majorana
masses to $\nu^c$ in $L^c$.
% One thing to note is that typical of SUSY models, we do have modulis along the
% matter and Higgs directions. For instance, 
% \begin{equation}
% S=0, \Phi_L=\Phi_L^c=0, \Phi_R=\Phi_R^c=M, H_{d1}=H_{u2}=0, H_{u1},H_{d2}\neq
% 0
% \end{equation}
% forms a moduli.
The missing partner mechanism pairs up $\Phi_L$ with $H_{d1}$ and $\Phi_L^c$
with $H_{u2}$. The charged components of $\Phi_R$ and $\Phi_R^c$ are either
eaten up or become sgoldstone partners of $W_R^{\pm}$, and some neutral
components are eaten up or become sgoldstone partners of $Z_R$. The spectrum is
summarized in Table \ref{tab:spec}.

\begin{table}
\[
\begin{array}{|c|c|}
\hline
\text{particle} 		& \text{mass}\\
\hline
S-\Phi_R^0 			& \sqrt{2}\kappa M\\
W_R^{\pm},\text{sgoldstone} 	& g_R M\\
Z_R,\text{sgoldstone} 		& \sqrt{g_R^2 + g_{B-L}^2}M\\
\Phi_L-H_{d1} 			& \alpha M\\
\Phi_L^c-H_{u2} 		& \beta M\\
H_{u1}, H_{d2} 			& 0\\
W_L^{\pm}, Z_L, \gamma 		& 0\\
\hline
\end{array}
\]
\caption{The Higgs supermultiplet and gauge spectrum of the minimal model at
the tree level in the limit where $M_{\frac{3}{2}}\to 0$.}
\label{tab:spec}
\end{table}

To get the seesaw mechanism to work, we need a $\Phi_R \Phi_R L^c L^c/\Lambda$
coupling. This can be obtained from a double seesaw mechanism if we introduce
an $SU(2)_L \times SU(2)_R \times U(1)_{B-L}$
singlet neutrino superfield $N$ and the couplings $\Phi_R
L^c N$ and
$N^2$. It appears that any additional symmetry which commutes with all the other
symmetries and allows all the couplings in Eq. \ref{eq:Wall} will also allow
couplings like $S^2$ and $S^3$. However, such couplings will not change our
analysis by much. Matter parity is still needed to forbid unwanted couplings.

% The low energy effective theory is gotten after integrating over the
% superheavy
% fields $S$, $\Phi_R$, the heavy gauge bosons, etc.

Let us assume for the moment that our model has $U(1)_{\mathcal{R}}$ symmetry
and see where our analysis leads us. $S$ has an
$\mathcal{R}$-charge of 2. By $C$-parity, $\Phi_L$ and $\Phi_R$, $\Phi_L^c$ and
$\Phi_R^c$, $Q$ and $Q^c$, and $L$ and $L^c$ will all have the same
$\mathcal{R}$-charge. $H_1$ and $H_2$
have to have the same $\mathcal{R}$-charge, and $\Phi_L$ and $\Phi_L^c$ opposite
$\mathcal{R}$-charges. Putting all of this together, we conclude that the
$\mathcal{R}$-charges of the $\Phi$ fields have to be zero and the
$\mathcal{R}$-charges of the bidoublets 2. The matter superfields have
zero $\mathcal{R}$-charges. We can also immediately see that matter parity has
to be a symmetry independent of $\mathcal{R}$-symmetry. This analysis nearly
works except that the seesaw term would now have to break
$\mathcal{R}$-symmetry. Because of this, we will not insist upon
$\mathcal{R}$-symmetry.

On the other hand, if we really want to have $\mathcal{R}$-symmetry, we may
consider using the mechanism proposed in Ref. \cite{Giudice:1988yz}.
Alternatively, instead of
the seesaw mechanism, we may arrange following Ref. \cite{Abel:2004tt} to have a
really tiny Dirac term. First,
we must forbid the direct coupling $H_1 L^c L$ even while we require the $H_2
L^c L$ coupling, which certainly demands some explanation. Instead, we have a
SUSY breaking sector involving a chiral superfield $X$ with an $R$-charge of $2$
with an intermediate scale SUSY breaking $F_X \simeq M_{3/2}\Lambda$ and a
K\"{a}hler term $X^\dagger H_1 L^c L/\Lambda$. This will induce a Dirac Yukawa
coupling of order $M_{3/2}/\Lambda$, where 
$\Lambda$ is several orders of
magnitude below the Planck scale.

We note that unlike other left-right models where $H_u$ and $H_d$ come
from the same bidoublet, our model does not necessarily predict CP-violation
because $H_u$ and $H_d$ 
come from different bidoublets and we can always perform a field redefinition
to make both $\alpha$ and $\beta$ real and positive.  Note however, that a tiny
VEV for $H_{d1}$ and $H_{u2}$ will be induced. These
VEVs are proportional to $M^{-2}$, which leads to negligible CP-violations
provided that $M$
 is large enough. The CP-violation coming
from the matter Yukawa sector remains, as it must.

Next, we consider the case where $SU(2)_R \times U(1)_{B-L}$ is broken at
some low energy scale. In Ref. \cite{Zhang:2007fn}, experimental bounds coming
from CP-violation as measured by neutral kaon oscillations place a bound of
about $2.5$ TeV on the mass of the $W_R^{\pm}$ gauge boson. See also Ref.
\cite{Wu:2007kt} which estimates a lower mass bound on $W_R^{\pm}$ of around
$600$ GeV and Ref. \cite{pdg} which gives a lower bound of $1.6$ TeV from the
kaon mass splitting.
From Table \ref{tab:spec}, with the dimensionless coefficients $\alpha$,
$\beta$ of order unity or less, we predict the existence of several new
particles which may be significantly
lighter than $W_R^{\pm}$ and accessible at the LHC.

When the $SU(2)_R$
breaking scale is high, the VEVs of $H_{d1}$ and $H_{u2}$
go to zero and we can redefine the phases of $H_1$ and $H_2$ independently so
that we do not have any contributions to CP-violation. However, as the breaking
scale goes down, these VEVs now become larger and more significant and they
contribute some amount to CP-violation from this sector. 

We also note
that the
spontaneous breaking of $C$-parity gives rise to $\mathbb{Z}_2$ domain walls. If
the symmetry breaking scale happens to lie below the reheating temperature, such
domain walls will appear after the universe has cooled down but they will
not be inflated away. This is potentially problematic from a cosmological point
of view.

There are at least two ways of giving $\nu^c$ (the neutral component of $L^c$)
a mass when $M$ is of the order of
a TeV or so; either through the double
seesaw mechanism or via a $SU(2)_L$ triplet. The former case has already been
discussed. We will now move on to the latter possibility, like that analyzed in
Ref. \cite{Dar:2005hm}. With such a low scale mass for $\nu^c$, we need to
suppress the Yukawa
Dirac
coupling by making it zero for instance, at the tree level. Radiative
corrections will then
generate a small value for the coupling resulting in the seesaw mechanism.
% Yet other possibility is that $\Phi$ happens to be a weak triplet instead of a
% weak doublet. This allows us to have a Dirac neutrino.
% $(\bm{1},\bm{3},\bm{1})_2$, $(\bm{1},\bm{3},\bm{1})_{-2}$,
% $(\bm{1},\bm{1},\bm{3})_{-2}$, $(\bm{1},\bm{1},\bm{3})_2$. With
% $\mathcal{R}$-symmetry the
% most general superpotential is given by
% \begin{equation}
% W = S f(T_L^c T_L+T_R^c T_R, (T_L^c T_L+T_R^c T_R)^2, T_L T_L T_R T_R, T_L^c
% T_L^c T_R^c T_R^c)
% \end{equation}
% assuming that the triplets have zero $\mathcal{R}$-charges. Otherwise, the
% most
% general
% coupling is given by
% \begin{equation}
% W = S f(T_L^c T_L+T_R^c T_R,(T_L^c T_L+T_R^c T_R)^2)
% \end{equation}
% A triplet breaking of $SU(2)_R$ leads to a residual $\mathbb{Z}_{2R}$
% gauge symmetry. In the former case, we do not have any accidental $U(1)_L$
% symmetry. In the latter, we do.
% 
% The soft SUSY breaking terms for the scalars are 
% \begin{equation}
% S, S^*, S^2, S^*S, c_1(T_L^*T_L+T_R^*T_R), c_2(T_L^{c*}T_L^c+T_R^{c*}T_R^c),
% c_3(T_L^c T_L+T_R^c T_R),..., c.c.
% \end{equation}
% $c_1$ and $c_2$ have to be real.
% 
% We have 26 real scalar fields. Four of the scalars, $S$ and some $T_R/T_R^c$
% linear combination get a mass from the superpotential. 2 of the goldstones are
% eaten up by $W'$, two become sgoldstones of $W'$, 1 is eaten up by $Z'$ and 1
% becomes a $Z'$ sgoldstone. We still have 16 light fields.

Let us now look at an alternative model with the $SU(2)_L$ triplets
$\Delta_L (\bm{1},\bm{3},\bm{1})_2$, $\Delta_L^c (\bm{1},\bm{3},\bm{1})_{-2}$,
$\Delta_R (\bm{1},\bm{1},\bm{3})_{-2}$ and
$\Delta_R^c (\bm{1},\bm{1},\bm{3})_2$. We introduce the complex conjugates to
cancel
the gauge anomalies, among other things. We will still need to have
Higgs bidoublets to give the matter Yukawa couplings at the renormalizable
level. To break the left-right mass relation, we still need the missing
partner mechanism which necessitates the presence of all the $\Phi$
fields\footnote{In addition, without the $\Phi$'s, we will still be left with
an unbroken $\mathbb{Z}_{2R}$ gauge symmetry.}. Actually, since we are going to
have so many additional particles above the $SU(2)_R$ breaking scale ($M$)
anyway, in the case where it happens to be low, it might not matter so much if
we have additional unpaired
superfields at the supersymmetric level since their soft masses would be
comparable to the masses that they would get from an $M$-scale pairing
anyway. Of course, it is also possible to consider a model where $M$ happens
to be large. But in that case, we need to find another model which reduces to
MSSM. This will be the second $\Delta$ model that we will consider. In
other words, this alternative model is really an extension of the model that
we have been studying previously. Because of this, $SU(2)_R$ will be broken by
 both $\Phi_R$ as well as $\Delta_R$. The primary reason for introducing these
 triplets is to give $\nu^c$ a mass via the coupling $\Delta_R L^c L^c$. 
 
%  This
% means that we definitely want $\Delta_R$ to get a significant VEV. This can be
%  arranged either by adding the coupling $S(\Delta_L^c \Delta_L + \Delta_R^c
% \Delta_R)$, or the coupling $\Delta_R^c \Phi_R \Phi_R$, $\Delta_R \Phi_R^c
% \Phi_R^c$
% and their $C$-conjugates. 
% With either choice, $\Delta_L$ and $\Delta_L^c$
% will remain light and we will not get MSSM at low energy scales. 

\begin{table}
\[
\begin{array}{||c|c||c|c||}
\hline\hline
\text{Superfield} 	& \text{Representation} 	& \text{Superfield} 
& \text{Representation}\\
\hline\hline
Q 			& (\bm{3},\bm{2},\bm{1})_{\frac{1}{3}} &
Q^c 			& (\bm{\overline{3}},\bm{1},\bm{2})_{-\frac{1}{3}}\\
\hline
L 			& (\bm{1},\bm{2},\bm{1})_{-1} &
L^c 			& (\bm{1},\bm{1},\bm{2})_1\\
\hline
H_1 			& (\bm{1},\bm{2},\bm{2})_0 &&\\
\hline
H_2 			& (\bm{1},\bm{2},\bm{2})_0 &&\\
\hline
S 			& (\bm{1},\bm{1},\bm{1})_0 &&\\
\hline
\Delta_L 		& (\bm{1},\bm{3},\bm{1})_2 &
\Delta_R 		& (\bm{1},\bm{1},\bm{3})_{-2}\\
\hline
\Delta_L^c 		& (\bm{1},\bm{3},\bm{1})_{-2} &
\Delta_R^c 		& (\bm{1},\bm{1},\bm{3})_2\\
\hline\hline
\end{array}
\]
\caption{The chiral superfield content of the reduced $\Delta$ model.}
\label{tab:reduced_triplet}
\end{table}

Let us first consider the minimal (reduced) $\Delta$ model which gives rise to
more
 low energy superfields than the MSSM. To break the up-down relation,
we still need two Higgs
bidoublets. This is more than what we have in MSSM, but all the additional
Higgs fields can be made massive by the soft SUSY terms.
If we assume that the
SUSY breaking scale is smaller than the $SU(2)_R$ breaking scale, then at the
LHC,
we would expect to see a doubly charged Dirac fermion and two doubly charged
scalars coming from $\Delta_R$ and $\Delta_R^c$, and a doubly charged Dirac
fermion and two doubly charged scalars coming from $\Delta_L$ and $\Delta_L^c$,
a charged Dirac fermion and two charged scalars coming from $\Delta_L$ and
$\Delta_L^c$, a neutral Dirac and four neutral scalars coming from $\Delta_L$
and $\Delta_L^c$, and two charged Dirac Higgsinos and two neutral Dirac
Higgsinos and three charged Higgs and seven neutral Higgs coming from the two
bidoublets.

Let us now turn to the second $\Delta$ model which reduces to MSSM if the
left-right symmetry breaking scale $M \gg$ TeV. While we will be primarily
interested in the case where $M$ is low, as this may give rise to new physics at
the LHC, this model will still be acceptable if $M$ is high.
A renormalizable
superpotential of the form
\begin{align}
W &\supset \kappa S (\Phi_L^c \Phi_L + \Phi_R^c \Phi_R + \rho
(\Delta_L^c\Delta_L + \Delta_R^c \Delta_R) - M^2) +
M'(\Delta_L^c\Delta_L + \Delta_R^c
\Delta_R) +\nonumber\\
& \quad \alpha (\Delta_R \Phi_R^c \Phi_R^c + 
\Delta_L \Phi_L^c \Phi_L^c) + \beta(\Delta_R^c \Phi_R
\Phi_R + \Delta_L^c \Phi_L \Phi_L)
\end{align}
will give rise to nonzero VEVs for both the $\Phi$'s and the $\Delta$'s. 
The equation $F_{\Delta_R}=F_{\Delta_R^c}=0$ causes $\langle \Delta \rangle$ to
be proportional to $\langle \Phi \rangle^2$. $F_S=0$ causes the VEVs to
be nonzero. Without any loss of generality, we may assume that there is no
$\Phi_L^c \Phi_L + \Phi_R^c \Phi_R$ term because any such term can always be
reabsorbed into a field redefinition of $S$ by some shift.

% If we are content to have light weak triplets, then we
% would expect that they would get masses from SUSY breaking and we might
% observe
% them at the LHC. In particular, we would expect to see particles with an
% electric charge of $\pm 2$. 

% With a bitriplet, we can construct a superpotential with 

To pair up the $\Delta_L$'s, we can introduce $K_L (\bm{3},\bm{2})_{-1}$, $K_R
(\bm{2},\bm{3})_1$, $K_L^c (\bm{3},\bm{2})_1$ and $K_R^c
(\bm{2},\bm{3})_{-1}$\footnote{Nonrenormalizable couplings like $\Phi_L^c
\Phi_L \Phi_R^c \Phi_R/\Lambda$ form an alternative, but for low scale
symmetry breaking, the cutoff scale $\Lambda$ would also have to be small.}.
Introduce the renormalizable couplings $\Delta_L K_L \Phi_R$, $\Delta_R K_R
\Phi_L$,
$\Delta_L^c K_L^c \Phi_R^c$ and $\Delta_R^c K_R^c \Phi_R^c$ and also the
couplings $K_L K_L \Delta_R^c$, $K_L^c K_L^c \Delta_R$ and their
$C$-conjugates. The $\Delta_L$'s pair up and the $\bm{3}_0$'s of the $K$'s also
pair up and all the masses are of the $\Lambda_R$ scale. A coupling like $K_L
K_L^c$ would be disastrous because the $\Delta_L$'s will only get seesaw
contributions to their masses. The full chiral superfield content of this model
is given in Table \ref{tab:triplet}.

\begin{table}
\[
\begin{array}{||c|c||c|c||}
\hline\hline
\text{Superfield} 	& \text{Representation} 	&
\text{Superfield} 
& \text{Representation}\\
\hline\hline
Q 			& (\bm{3},\bm{2},\bm{1})_{\frac{1}{3}} &
Q^c 			& (\bm{\overline{3}},\bm{1},\bm{2})_{-\frac{1}{3}}\\
\hline
L 			& (\bm{1},\bm{2},\bm{1})_{-1} &
L^c 			& (\bm{1},\bm{1},\bm{2})_1\\
\hline
H_1 			& (\bm{1},\bm{2},\bm{2})_0 &&\\
\hline
H_2 			& (\bm{1},\bm{2},\bm{2})_0 &&\\
\hline
S 			& (\bm{1},\bm{1},\bm{1})_0 &&\\
\hline
\Phi_L 			& (\bm{1},\bm{2},\bm{1})_1 &
\Phi_R 			& (\bm{1},\bm{1},\bm{2})_{-1}\\
\hline
\Phi_L^c 		& (\bm{1},\bm{2},\bm{1})_{-1} &
\Phi_R^c 		& (\bm{1},\bm{1},\bm{2})_1\\
\hline
\Delta_L 		& (\bm{1},\bm{3},\bm{1})_2 &
\Delta_R 		& (\bm{1},\bm{1},\bm{3})_{-2}\\
\hline
\Delta_L^c 		& (\bm{1},\bm{3},\bm{1})_{-2} &
\Delta_R^c 		& (\bm{1},\bm{1},\bm{3})_2\\
\hline
A_L 			& (\bm{1},\bm{3},\bm{1})_0 &
A_R 			& (\bm{1},\bm{1},\bm{3})_0\\
\hline
K_L 			& (\bm{1},\bm{3},\bm{2})_{-1} &
K_R 			& (\bm{1},\bm{2},\bm{3})_1\\
\hline
K_L^c 			& (\bm{1},\bm{3},\bm{2})_1 &
K_R^c 			& (\bm{1},\bm{2},\bm{3})_{-1}\\
\hline\hline
\end{array}
\]
\caption{The chiral superfield content of the $\Delta$ model which reduces to
MSSM at low energies.}
\label{tab:triplet}
\end{table}

We also need to give masses to the doubly-charged component of $\Delta_R$.
As already mentioned, a direct $\Delta_R^c \Delta_R$ will not do. So, let us
introduce an additional $A_L (\bm{3},\bm{1})_0$ and $A_R (\bm{1},\bm{3})_0$ and
the couplings $\Delta_L \Delta_L^c A_L$ and $\Delta_R \Delta_R^c A_R$. The
last term in addition to the mass term $A_R^2$ (and its $C$-conjugate) will 
induce a nonzero VEV ($\sim M$) for $A_R$ since $\Delta_R$ and $\Delta_R^c$
already
have nonzero VEVs. With so
many additional light particles at the TeV scale, we will have plenty of new
physics to work with at the LHC. 

In summary, the superpotential will contain the following renormalizable terms:
\begin{align}
W &\supset S, S(\Phi_L^c \Phi_L + \Phi_R^c \Phi_R), (\Delta_L^c \Delta_L +
\Delta_R^c \Delta_R),\nonumber\\
 &\quad (\Delta_L\Phi_L^c\Phi_L^c +
\Delta_R\Phi_R^c\Phi_R^c),
(\Delta_L^c\Phi_L\Phi_L + \Delta_R^c \Phi_R\Phi_R), (A_L\Phi_L^c\Phi_L + A_R
\Phi_R^c\Phi_R^c), A_L^2, A_R^2,\nonumber\\
&\quad 
 H_1\Phi_L\Phi_R,
H_2\Phi_L^c\Phi_R^c,\nonumber\\
&\quad K_L\Delta_L\Phi_R+K_R\Delta_R\Phi_L,
K_L^c\Delta_L^c\Phi_R^c+K_R^c\Delta_R^c\Phi_L^c,K_L^c K_L, K_R^cK_R.
\end{align}

To sum up, we predict several
$SU(2)_L$ Higgs triplets, $SU(2)_L$ Higgs doublets, and $SU(2)_L$
singlets at low
energies. 
% The 1-loop $\beta$-function coefficient for $\alpha_L$ is now $-20$, which is
% so large that $\alpha_L$ will rise so steeply that we will hit a Landau pole
% around $10^6$ GeV to $10^7$ GeV. 
% This is a really low scale. However, if
% $SU(2)_L$ and $SU(2)_R$ are unified in some larger gauge group below that
% scale, it might be possible to change the $\beta$ function before then to
% avoid
% the Landau pole. This is because the contribution to the $\beta$-function
% coming from vector multiplets is of the opposite sign. In addition, a vector
% multiplet contribution is three times that of a chiral multiplet contribution.
The $\Delta_L$'s and both $K_L$ and $K_R$ will
contribute doubly charged particles. In particular, we will have six doubly
charged scalars and three doubly charged Dirac
particles. These particles all have masses comparable to or smaller than the
mass of $W_R^{\pm}$.

\section{Conclusion}
% In conclusion, we note that previous analyses of left-right models have an
% unwanted left-right mass relation which we correct for using the missing
% partner mechanism. This mechanism has the additional benefit of getting rid of
% additional light states to the MSSM so that we are left with pure MSSM at low
% energies. With low scale $SU(2)_R$ breaking, we can introduce weak triplets to
% implement the seesaw mechanism, but this will lead to a lot of additional
% light
% states and the gauge coupling strength will blow up at some low scale. This
% kind of argues against a low energy breaking of our models. 

In this paper we have insisted that the scalar (Higgs) sector
of models based on symmetry groups such as $SU(2)_L \times SU(2)_R \times
U(1)_{B-L}$
should respect $C$-parity.
We have
considered a variety of models to show how the 
MSSM can be recovered at energies below the left-right ($C$-parity) 
symmetry breaking scale. If the latter
happens to lie in the TeV range a plethora of new particles should be
accessible at the LHC.

\begin{acknowledgments}
We thank Ilia
Gogoladze for helpful discussions. This work is supported by DOE Grant No.
DE-FG02-91ER40626.
\end{acknowledgments}

\end{document}